\documentclass[12pt]{article}
\usepackage{times}
\usepackage{graphicx}
\usepackage{amsmath,amssymb}
\usepackage[utf8]{inputenc}
\usepackage{textcomp}
\usepackage{gensymb}
\usepackage[version=4]{mhchem}
\usepackage{siunitx}
\usepackage{longtable,tabularx}
\usepackage{hyperref}
\usepackage[numbers]{natbib}
\usepackage[a4paper, margin=1in]{geometry}

\title{Radiometric Interferometry for Deep Space Navigation using Geostationary Satellites}

\author{Moshe Golani, Yoram Rozen, Hector Rotstein\\
Technion, Israel Institute of Technology, Haifa, Israel\\
\texttt{moshe.golani@campus.technion.ac.il}}

\date{}

\begin{document}
\maketitle
\begin{abstract} Deep space navigation - defined as spacecraft tracking beyond the lunar orbit - presents significant challenges due to the unavailability of Global Navigation Satellite System (GNSS) signals and severe signal attenuation over interplanetary distances. Traditional terrestrial systems, such as NASA’s Deep Space Network (DSN) and ESA’s ESTRACK, rely on Very Long Baseline Interferometry (VLBI) for angular positioning. However, these systems are limited by relatively short baselines, atmospheric distortions requiring extensive calibration, and reduced Line-of-Sight (LOS) availability due to Earth’s rotation. Because VLBI angle measurements require at least two simultaneously visible stations, the measurement duty cycle is inherently constrained.

This research proposes a complementary deep space navigation approach using space-based interferometry, in which radio signals from the spacecraft are received and cross-correlated onboard Geostationary Earth Orbit (GEO) satellites. By replacing terrestrial VLBI stations with dual GEO platforms, the method significantly extends the effective baseline, removes atmospheric phase errors, and provides near-continuous visibility to deep space targets. Unlike Earth-based systems, GEO-based interferometry maintains persistent station mutual visibility, enabling higher measurement availability and more flexible mission support.

A complete system model is presented, including the principles of dual-frequency phase-based angular tracking and a structured error budget analysis. Theoretical results, combined with detailed simulations, show that the GEO-based system achieves a total angular error of approximately 3.73 nanoradians, within the same order of magnitude as terrestrial VLBI. In particular, the space-based architecture nearly doubles the geometrical availability for interferometric tracking, while eliminating atmospheric calibration requirements. These findings support the feasibility of the GEO-based VLBI concept and motivate continued research and field validation for future deep-space navigation applications.
\end{abstract}


\newpage
\tableofcontents
\newpage

\section{Introduction}

Deep space navigation presents significant challenges, particularly for small spacecraft with limited transmission power and onboard resources \cite{turan2022autonomous, vittorio2021autonomous}. Traditional Global Navigation Satellite Systems (GNSS), widely used for orbit determination (OD) in Earth-orbiting satellites, become ineffective beyond distances of a few tens of thousands of kilometers \cite{ParkWonKwon2017}. The primary limitation is the severe power attenuation of GNSS signals, which follows a free-space loss (FSL) proportional to $1/R^2$. For instance, an electromagnetic signal propagating over a distance of 1 million kilometers experiences a power density loss of approximately $190$ dB. Additionally, unfavorable triangulation geometry results in a narrow vertex angle and a high Geometric Dilution of Precision (GDOP), further degrading the accuracy of GNSS-based navigation solutions in deep space.

As a result, alternative navigation methods are required. Optical navigation \cite{andreis2021overview} and radiometric interferometry, which utilizes phase measurements of electromagnetic waves \cite{book2013delta, curkendall2013delta}, are commonly used techniques for deep space orbit determination \cite{ely2022comparison}. The most established radiometric approach is NASA’s Deep Space Network (DSN) \cite{mudgway2001uplink} and its European counterpart, ESA’s ESTRACK \cite{doat2018esa}, both of which rely on Very Long Baseline Interferometry (VLBI) performed by ground-based stations. The DSN, in particular, consists of globally distributed antenna arrays, with dish sizes up to 70 meters in diameter, designed to support deep space communications and navigation \cite{layland1997evolution}. While these terrestrial systems have demonstrated exceptional accuracy, they are fundamentally constrained by their ground-based nature.

The primary limitations of DSN and ESTRACK in deep space navigation include:
\begin{itemize}
    \item \textbf{Short baselines in deep space scenarios:} Although VLBI accuracy improves with increasing baseline length, the maximum baseline achievable by ground-based stations is limited to Earth's diameter ($\sim 12,700$ km). Terrestrial systems baseline is typically $\sim 8,000$ km.  This constraint limits the angular resolution for deep space targets.
    \item \textbf{Atmospheric distortions:} Ionospheric and tropospheric effects introduce phase errors that require extensive calibration, adding complexity to the measurement processing and potentially degrading the final accuracy.

\item \textbf{Limited simultaneous station visibility:} VLBI requires that at least two stations have concurrent line-of-sight to the target spacecraft. Due to Earth’s curvature and rotation, this condition is satisfied only about 50\% of the time during a diurnal cycle. As a result, observation windows are inherently limited and fragmented, which reduces the overall availability of angular measurements opportunities for navigation. 

\item \textbf{High demand and limited access:} DSN and ESTRACK assets are heavily utilized for mission-critical operations across multiple agencies. As a result, obtaining sufficient observation time on these networks is increasingly difficult, posing a major barrier for navigation support of additional missions.

\end{itemize}

To overcome these constraints, this research proposes an alternative deep space navigation approach using space-based radio interferometry. Specifically, this method replaces terrestrial VLBI stations with phase sensors onboard Geostationary Earth Orbit (GEO) satellites. This configuration offers several key advantages:
\begin{itemize}
    \item \textbf{Extended baselines:} The separation between GEO satellites can exceed 80,000 km - an order of magnitude greater than the terrestrial systems limited by Earth's diameter - creating substantial potential for improving angular measurement precision.
    \item \textbf{Elimination of atmospheric distortions:} Since GEO-based interferometry operates in space, it avoids ionospheric and tropospheric distortions, eliminating the need for complex atmospheric calibration.
    \item \textbf{Higher system availability:} Unlike ground stations, GEO satellites maintain near-continuous line-of-sight (LOS) to deep space targets, removing Earth-based LOS constraints and ensuring more frequent angular measurements.
    \item \textbf{Independent space-based navigation infrastructure:} By reducing reliance on Earth-based VLBI networks, this method enables a more autonomous and scalable deep space navigation system.
\end{itemize}

This paper is structured as follows: The next section outlines the fundamental principles of terrestrial VLBI and its limitations for deep space navigation. Section \ref{proposed method section} introduces the proposed space-based VLBI system, detailing its measurement methodology, system architecture, and key design considerations.
Section \ref{performance analysis} discusses the error individual and aggregated contributions and the resulted expected performance. Section \ref{simulation} presents preliminary simulation results demonstrating the feasibility of GEO-based phase sensors, analyzing system availability improvements and phase noise characteristics. Finally, Section \ref{conclusions} summarizes the findings and discusses future research directions.

\section{Fundamentals of VLBI for Deep Space Navigation}

Very Long Baseline Interferometry (VLBI) is a well-established technique originally developed in the latter half of the 20\textsuperscript{th} century for applications in radio astronomy and geodesy \cite{schuh2012vlbi}, and later extended to satellite orbit determination. Traditional and more recent applications of VLBI in satellite tracking are well documented, for instance, in \cite{hellerschmied2018satellite}.

In the context of deep space navigation, VLBI is employed to determine the angular position of a distant spacecraft relative to a known terrestrial baseline \cite{fiori2022deep}. This section outlines the fundamental principles and derivations used in terrestrial VLBI systems to extract such angular information. Since the proposed space-based method builds upon the same core principles, this section provides a technical foundation for comparison with the RINGS concept introduced later in Section~\ref{proposed method section}. The contrast will serve to highlight the advantages and novel capabilities of the GEO-based interferometric approach.

\subsection{Angle Measurement by VLBI}
VLBI relies on two widely separated high-gain receiving antennas that capture and record the spacecraft's signal within a specific frequency band. These signals are subsequently processed using a software correlator to extract the Differential One-way Ranging (DOR) data between the antennas, enabling precise determination of the spacecraft's direction. The principles of VLBI-based direction finding are thoroughly discussed in \cite{book2013delta, fiori2022deep}.

\begin{figure}[hbt!]
\centering
\includegraphics[width=0.7\textwidth]{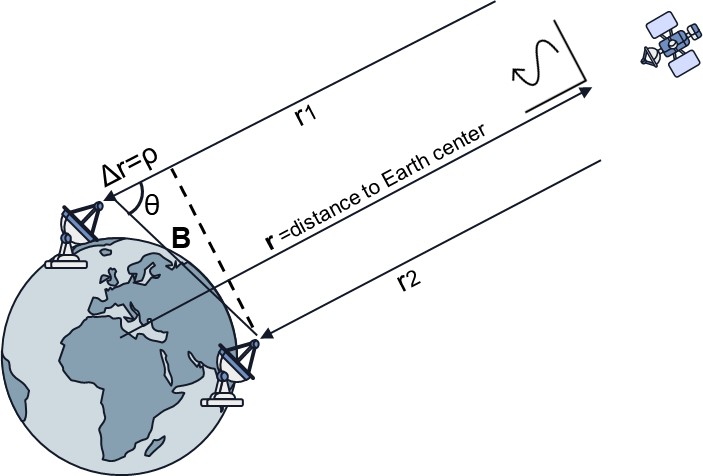}
\caption{Spacecraft angular position determination by Differential One-Way Ranging (DOR) using Earth-based VLBI stations.}
\label{fig:Basic DOR}
\end{figure}

As illustrated in Fig. \ref{fig:Basic DOR}, DOR measurements are taken simultaneously by two Earth-based stations at precisely known locations. The spacecraft's direction of arrival (DOA) relative to the baseline is denoted by $\theta$ and is given by:

\begin{equation} \label{DOR1}
  \Delta r = c\, \Delta{t}=B \cos {\theta}  
\end{equation} 
\begin{equation} \label{DOR1.0}
    \cos{\theta} = \frac {c\, \Delta{t}}{B}  
\end{equation} 
\begin{equation} \label{DOR1.1}
  \theta=\arccos {\frac{c\, \Delta{t}}{B}}  
\end{equation} 

where $B$ is the baseline vector connecting the two receivers, $\Delta r$ is the measured range difference in the direction of the spacecraft, $\Delta t$ is the measured time difference, and $c$ is the speed of light.

\paragraph{Measurement Sensitivity} The sensitivity of the estimated $\theta$ to time difference errors is derived by differentiating equation \ref{DOR1.0} with respect to $\Delta t$ \cite{fiori2022deep}:

\begin{equation}\label{DOR2}
    \frac{d\theta}{d\Delta t}=-\frac{c}{B \sin{\theta}}.
\end{equation}

For small angles, $\sin{\theta}$ can be approximated using the first-order terms of its Taylor expansion, $\sin{\theta}\approx{\theta-\theta^3/6}$, leading to:

\begin{equation}\label{DOR3}
    \frac{d\theta}{d\Delta t}=-\frac{c}{B}\frac{1}{\theta \left( 1-\frac{\theta^2}{6}\right)}.
\end{equation}

This expression shows that the sensitivity of $\theta$ to errors in $\Delta t$ is inversely proportional to the baseline length $B$ and the angle $\theta$. Since $\theta$ is bounded within $[0, \pi]$, and the singularity at $\theta=0$ is generally avoided in practice, the dominant factor in reducing sensitivity to errors is increasing $B$.

\subsection{Electromagnetic Phase Measurement in VLBI}
The fundamental measurement quantity in a VLBI system is the electromagnetic wave phase received by each ground station antenna. The far-field solution for a planar electromagnetic wave is given by:

\begin{equation}\label{E1}
    E_{i}(t,r)=E_{0}\,e^{j(\overrightarrow{k}\cdot\overrightarrow{r}-\omega t)}
\end{equation}

where:
\begin{itemize}
    \item $E_{0}$ is the amplitude of the electric field,
    \item $\overrightarrow{k}$ is the wave vector with magnitude $2\pi/\lambda$,
    \item $\overrightarrow{r}$ is the propagation distance,
    \item $\omega = 2\pi f_{c}$ is the angular frequency of the wave, and
    \item $t$ is the time coordinate.
\end{itemize}

The phase information, which contains the key measurement data for interferometry, is extracted from the exponent:

\begin{equation}
    \phi_{i}=\overrightarrow{k}\cdot\overrightarrow{r_{i}}-\omega t +\phi_{0}.
\end{equation}

By computing the phase difference between two ground-based receivers (Station 1 and Station 2), we obtain:

\begin{equation}\label{DPHI}
    \Delta\phi=\frac{2\pi}{\lambda}\,r_{1}-\frac{2\pi}{\lambda}\,r_{2}=\Delta r \,\frac{2\pi}{\lambda}.
\end{equation}

Substituting equation \ref{DOR1}:

\begin{equation} \label{difi}
    \Delta\phi=\frac{2\pi}{\lambda}\,B \cos {\theta}.
\end{equation}

Solving for $\theta$:

\begin{equation}
    \cos{\theta} = \frac{\Delta\phi\,\lambda} {2\pi B}=\frac{c}{B}\,\frac{\Delta\phi}{2\pi f}.
\end{equation}

\subsection{Ambiguity Resolution in VLBI Phase Measurements}
One of the major challenges in phase-based interferometry is phase ambiguity, which arises because phase measurements are inherently modulo $2\pi$. The measured phase difference $\Delta\phi$ alone does not provide the integer number of full wavelengths between receivers. 

To resolve this ambiguity, multiple site locations, multiple frequencies, or a combination of both can be used \cite{book2013delta, curkendall2013delta}. The dual-frequency approach provides an effective solution by computing the group delay:

\begin{equation} \label{DDFI}
 \Delta\Delta\phi = B \cos{\theta} \left(\frac{2\pi}{\lambda_{1}}-\frac{2\pi}{\lambda_{2}}\right)=2\pi\frac{B}{c}\cos{\theta}\,(f_{1}-f_{2}).
\end{equation}

Rearranging:

\begin{equation}\label{cosTheta}
 \cos{\theta}=\frac{c}{B}\,\frac{\Delta\Delta\phi}{2\pi\Delta f_{c}}.
\end{equation}

For a dual-frequency system, the final ambiguity-free angular measurement is:

\begin{equation} \label{final theta}
  \theta=\arccos \left(\frac{c}{B}\,\frac{\Delta\Delta\phi}{2\pi\Delta f_{c}}\right).
\end{equation}

This technique dramatically reduces ambiguity effects, shifting the uncertainty from the scale of the carrier wavelength $\lambda$ to the much larger synthetic wavelength $c/\Delta f_{c}$, enabling robust and precise deep-space angle measurements.

\subsection{Doppler Effect Analysis in Terrestrial VLBI}

In VLBI systems, the Doppler effect introduces a frequency shift in the received signal due to the relative motion between the spacecraft and each receiving station. For Earth-based systems, this relative motion is primarily governed by Earth's rotation, which imparts differing velocities to spatially separated antennas.

The received frequency $f_{d}$, including special relativity effects, is given by:

\begin{equation} \label{rel-dop}
    f_{d} = f_{0} \frac{\sqrt{1 - v^2/c^2}}{1 - v_{r}(t)/c}
\end{equation}

where $f_{d}$ and $f_{0}$ are the received and transmitted frequencies, respectively; $v$ is the station's total velocity in an inertial frame; $v_{r}(t)$ is the radial component of the velocity vector, defined as the projection of the station's velocity along the line-of-sight to the spacecraft; and $c$ is the speed of light.

Since $v^2/c^2 \ll 1$ for both spacecraft and stations (typically on the order of $10^{-10}$), relativistic corrections can be neglected, yielding the classical Doppler shift:

\begin{equation} \label{fd}
    f_{d} = f_{0} \left(1 + \frac{v_{r}(t)}{c}\right)
\end{equation}

The corresponding frequency offset, defined by $f_{d} - f_{0}$, is:

\begin{equation} \label{fdop}
    \Delta f_{d} = f_{0} \frac{v_{r}(t)}{c} = \frac{v_{r}(t)}{\lambda_{0}}
\end{equation}

Since VLBI measures signal phase, the Doppler effect modifies the effective wavelength at each station:

\begin{equation}
    \lambda_{d} = \frac{c}{f_{d}} = \frac{c}{f_{i}\left(1 + \frac{v_{r_j}}{c}\right)},
\end{equation}

where $f_i$ denotes the signal carrier frequency, and $j = 1,2$ indexes the receiving stations. The received phase at station $j$ is given by:

\begin{equation}
    \phi_{f_{i,j}} = \frac{2\pi f_i}{c} \left(1 + \frac{v_{r_j}}{c}\right) r_j,
\end{equation}

with $r_j$ being the distance from station $j$ to the spacecraft and $v_{r_j}$ its radial velocity component.

The phase difference between the two stations at frequency $f_1$ is:

\begin{equation}
    \Delta \phi_{f_1} = \frac{2\pi f_1}{c} \left[ \left(1 + \frac{v_{r_1}}{c}\right) r_1 - \left(1 + \frac{v_{r_2}}{c}\right) r_2 \right],
\end{equation}

and similarly for $f_2$:

\begin{equation}
    \Delta \phi_{f_2} = \frac{2\pi f_2}{c} \left[ \left(1 + \frac{v_{r_1}}{c}\right) r_1 - \left(1 + \frac{v_{r_2}}{c}\right) r_2 \right].
\end{equation}

Subtracting these gives the dual-frequency phase difference:

\begin{equation}
    \Delta\Delta\phi = \Delta \phi_{f_1} - \Delta \phi_{f_2} = \frac{2\pi}{c} \Delta f_c \left[ \left(1 + \frac{v_{r_1}}{c}\right) r_1 - \left(1 + \frac{v_{r_2}}{c}\right) r_2 \right],
\end{equation}

where $\Delta f_c = f_1 - f_2$ is the frequency separation.

We define the baseline projection as $r_1 - r_2 = B \cos\theta$, where $B$ is the baseline length and $\theta$ is the angle between the baseline vector and the line-of-sight to the source, as illustrated in Fig. \ref{fig:Basic DOR}. For the involving the products $r_1 v_{r_1}$ and $r_2 v_{r_2}$, we can assume that $r_1 \approx r_2 \approx r$, maintaining $r_1 + r_2 = 2r$, therefore the expression simplifies to:

\begin{align}
    \Delta\Delta\phi &= \frac{2\pi \Delta f_c}{c} \left[ B \cos\theta + \frac{r}{c} \left(v_{r_1} - v_{r_2}\right) \right]
\end{align}

We now define the vectorial radial velocity difference as:

\begin{equation}
    \Delta V_r \triangleq \mathbf{V}_{r_1} \cdot \hat{\mathbf{s}} - \mathbf{V}_{r_2} \cdot \hat{\mathbf{s}} = v_{r_1} - v_{r_2},
\end{equation}

where $\hat{\mathbf{s}}$ is the unit vector pointing from the observer toward the spacecraft, and $\mathbf{V}_{r_j}$ are the velocity vectors of the stations. Then:

\begin{equation} \label{DDphi_dop_vector}
    \Delta\Delta\phi = \frac{2\pi \Delta f_c}{c} \left[ B \cos\theta + \frac{r}{c} \Delta V_r \right]
\end{equation}

This general expression shows that the differential phase $\Delta\Delta\phi$ includes two additive components:
\begin{itemize}
    \item An undistorted \textit{geometric term}, identical to the basic equation \ref{DDFI}, proportional to $B \cos\theta$, capturing the baseline projection along the source direction.
    \item A \textit{Doppler-induced correction term} proportional to $r \, \Delta V_r / c$, which reflects the differential motion of the receiving stations along the line-of-sight.
\end{itemize}

While Doppler contributions may not be the dominant term in interferometric phase measurements, they are non-negligible and must be explicitly modeled to ensure accurate interpretation of the data. In terrestrial VLBI systems, the differential Doppler effect is typically smaller, primarily due to the modest tangential velocities introduced by Earth’s rotation. Nevertheless, for high-precision angular estimation, even these small Doppler components require correction to avoid systematic bias.

In space-based VLBI configurations - such as those involving GEO satellites or other high-velocity platforms - the differential radial velocity, denoted $\Delta V_r$, between observing stations can become a significant contributor to the measured phase. Proper modeling of this effect is essential to achieve measurement accuracy.

This general Doppler formulation will be later applied to the specific case of a symmetric GEO-based VLBI geometry.

\section{Proposed System: Radiometric Interferometry Navigation by GEO Satellites (``RINGS'')} \label{proposed method section}

\begin{figure}[hbt!]
\centering
\includegraphics[width=0.99\linewidth]{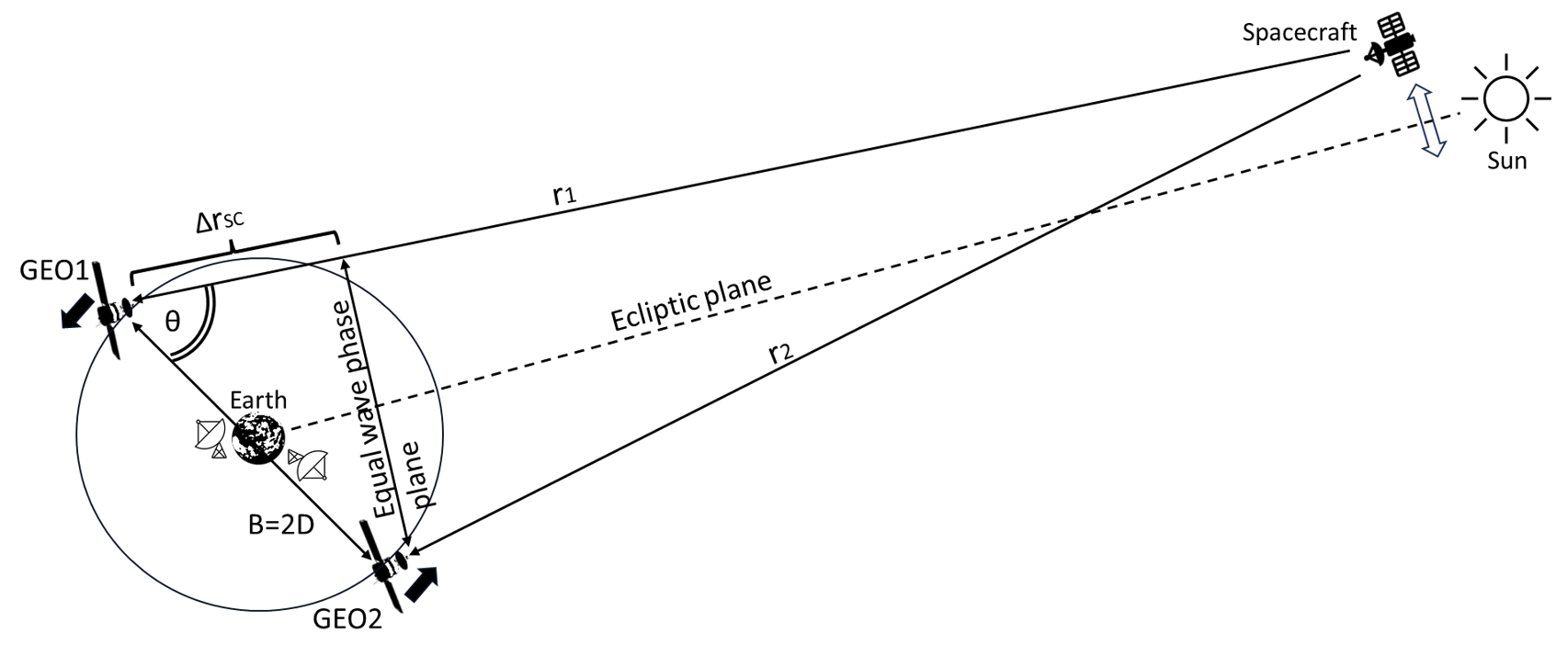}
\caption{Space-based Dual-GEO satellite DOR array. $\Delta{r_{sc}}$ is the projected range difference between the satellites with respect to signal direction of arrival.}
\label{fig:Basic RINGS}
\end{figure}

\begin{figure}[hbt!]
\centering
\includegraphics[width=1.0\linewidth]{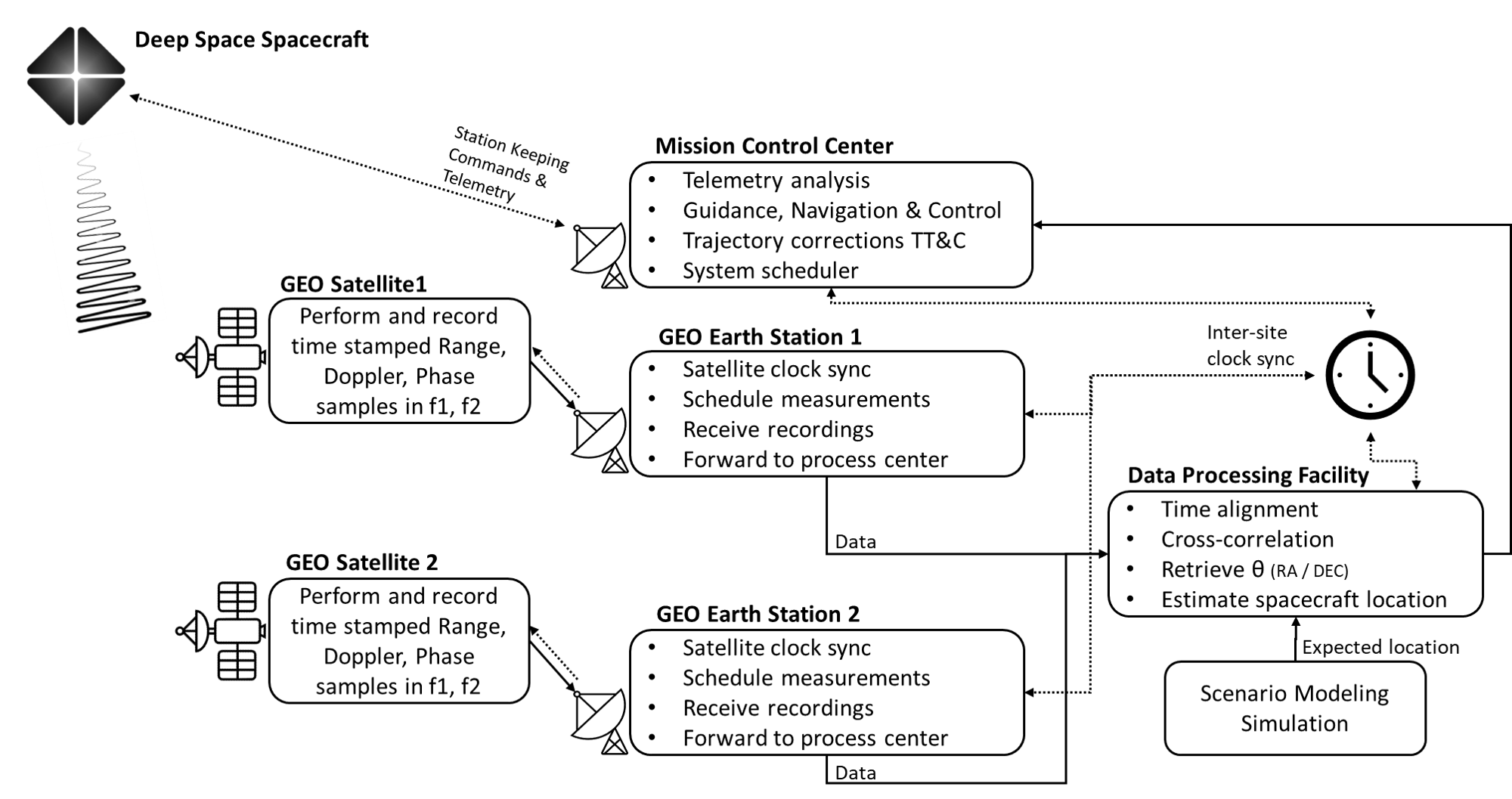}
\caption{General system block diagram for the GEO satellite based VLBI (RINGS).}
\label{fig:block diagram}
\end{figure}

This section outlines the proposed method including a general system architecture, principles of operation, and data processing approach used to obtain angular measurements of a spacecraft's position in space. We propose a new VLBI configuration that replaces terrestrial baselines with two geostationary (GEO) satellites positioned at widely separated longitudes. This configuration forms a space-based interferometric baseline exceeding that of any ground-based system, as illustrated in Fig.~\ref{fig:Basic RINGS}. A high-level functional diagram of the system main components and and the general flow is shown in Fig.~\ref{fig:block diagram}.

During a measurement session lasting approximately 2 to 5 minutes, the target spacecraft’s radio signal is simultaneously received by both GEO satellites, time-stamped with high precision, and stored onboard. These observations are subsequently downloaded via two GEO-compatible ground stations and forwarded to a central processing facility. There, the signals are time-aligned and cross-correlated to extract phase differences, which are then used for precise angular localization. The derived navigation solution is delivered to the Mission Control Center to support real-time Guidance and Control (G\&C) operations, including orbit correction commands.

The next section examine the primary contributors to interferometric error in the proposed system: signal-to-noise ratio, Doppler-induced phase bias, clock synchronization and location uncertainties, and overall system-level error budget. 

\section{Performance Analysis Discussion} \label{performance analysis}

This section applies the structured error estimation methodology developed and used by NASA and ESA \cite{iess2014astra, spencer2023interplanetary} for terrestrial VLBI and Delta-DOR systems to the proposed GEO-based VLBI concept. Following the methodology introduced and elaborated by various authors \cite{becker2013autonomous, curkendall2013delta, book2013delta, fiori2022deep, iess2014astra}, we adopt the same framework to evaluate the dominant error sources contributing to the total phase and angular uncertainty in the RINGS system.

This approach ensures consistency with established ground-based VLBI error budgets, facilitating direct comparison between terrestrial and space-based configurations. The analysis identifies the critical factors limiting system performance and highlights areas where the space-based VLBI architecture offers inherent advantages. Key error contributors include signal-to-noise ratio limitations, Doppler-induced phase shifts, clock synchronization uncertainty, and orbit determination errors of the GEO satellites. Each component is examined to derive a comprehensive error budget that supports system-level performance predictions.

\subsection{Evaluation of the Signal to Noise (SNR)}

A primary contributor to the error budget in the satellite-based VLBI configuration is the relatively low received signal power from deep-space spacecraft. This, combined with systematic noise sources, results in a modest signal-to-noise ratio (SNR), which directly impacts the accuracy of phase extraction in the interferometric process.

The total system noise arises primarily from two independent sources:

\begin{enumerate}
    \item \textbf{Thermal noise} - determined by the receiver's ambient noise temperature and the cosmic background radiation incident on the antenna aperture.
    \item \textbf{Phase noise} - originating from imperfections in the onboard oscillators and stochastic propagation effects such as scintillation and plasma-induced variations.
\end{enumerate}

In this section, we quantify the contribution of each noise source and compute the resulting effective system noise. We then derive the received signal power based on standard link budget formulations. Finally, the SNR is obtained by comparing the received signal power to the combined noise power, establishing a baseline for subsequent phase error and angular uncertainty analysis.

\subsubsection{Thermal Noise Power}

The total system noise temperature is given by the sum of the receiver's ambient noise and the cosmic background noise received by the antenna aperture. Unlike terrestrial VLBI systems, there is no atmospheric absorption that attenuates the incoming signal, nor is there atmospheric emission that adds to the system temperature, as commonly accounted for in ground-based models \cite{felli2012very}. Contributions from planetary bodies and the Sun’s blackbody radiation are neglected due to their relatively small angular coverage compared to the antenna’s field of view. Therefore,

\begin{equation}\label{eq:T_op}
   T_{op} = T_{amb} + T_{sky}
\end{equation}

where \( T_{amb} \approx 290~\text{K} \) is the receiver’s ambient noise temperature, and \( T_{sky} = 2.7~\text{K} \) represents the cosmic microwave background observed by the antenna aperture. Thus,

\begin{equation}\label{eq:T_op_value}
   T_{op} \approx 290 + 2.7 = 292.7~\text{K}
\end{equation}

The thermal noise power at the receiver input is given by the Johnson–Nyquist equation \cite{turner2012johnson}:

\begin{equation}\label{eq:NoisePower}
   N_{\text{T}} = k_{B}\, T_{op}\, BW
\end{equation}

where \( k_{B} = 1.38 \times 10^{-23}~\text{J/K} \) is Boltzmann’s constant, and $BW$ is the receiver bandwidth - a variable system parameter, temporarily set to \(1~\text{kHz} \). Substituting the values:

\begin{equation}\label{eq:NoisePower_value}
   N_{\text{T}} = (1.38 \times 10^{-23}) \times (292.7) \times (10^3) \approx 4.04 \times 10^{-18}~\text{W} \quad (-174~\text{dBW})
\end{equation}

\subsubsection{Phase Noise Contribution}

Phase noise arises from two primary sources:

\begin{itemize}
    \item \textbf{Internal phase noise:} Generated by oscillator instabilities in both the transmitter and receiver. Since the same transmitted signal is received at both GEO satellites, any static or slowly varying transmitter phase noise is effectively canceled in the differential measurement. However, rapid fluctuations or time-dependent variations may still introduce residual errors.
    
    \item \textbf{External phase noise:} Introduced by the propagation medium, primarily due to interplanetary plasma fluctuations, scintillations, and other stochastic effects along the wavefront. These effects are not common to both receivers and thus contribute directly to the phase measurement uncertainty.
\end{itemize}

For the purpose of a preliminary noise budget, we adopt a conservative upper bound for the total phase noise variance. This serves as a rough order-of-magnitude estimate and will later be refined based on the actual signal-to-noise ratio (SNR) and expected dispersion conditions. In practice, phase noise variance should be derived from oscillator stability specifications and propagation models, or inferred from historical data.

We therefore assume the accumulated phase uncertainty follows a zero-mean Gaussian distribution with variance \cite{felli2012very}:

\begin{equation}\label{eq:phase_noise_variance}
    \sigma_{\phi}^2 \approx 10^{-2}~\text{rad}^2.
\end{equation}

\subsubsection{Effective Noise Power Including Phase Noise}
When both thermal noise and phase noise are significant, the total effective noise power is given by:

\begin{equation}\label{eq:total_noise_power}
   N_{\text{RF}} = \sqrt{N_{\text{T}}^2 + N_{\text{PH}}^2}.
\end{equation}

Here, $N_{T}$ refers to the thermal noise power and $N_{PH}$ - the phase noise power contribution is defined as:

\begin{equation}\label{eq:phase_noise_power}
   N_{\text{PH}} = \sigma_{\phi}^2.
\end{equation}

Therefore, the effective SNR for a single interferometric phase sample is given by:

\begin{equation}\label{eq:SNReff}
   \text{SNR}_{\text{eff}} = \frac{S}{\sqrt{N_{\text{T}}^2 + N_{\text{PH}}^2}},
\end{equation}

where $S$ is the received signal power, $N_{\text{T}}$ is the thermal noise power, and $N_{\text{PH}}$ is the equivalent phase noise contribution, converted to power units.

For low phase noise variance (\(\sigma_{\phi}^2 \ll 1\)), the impact of phase noise on overall SNR is typically minor. However, if phase noise increases beyond a certain threshold, it may significantly degrade the system's ability to extract accurate phase measurements.

Since the phase noise contribution is typically small compared to the thermal noise power in deep-space scenarios, the expression can be approximated as:

\begin{equation}\label{eq:SNR_simple}
   \text{SNR}_{\text{eff}} \approx \frac{S}{N_{\text{T}}}.
\end{equation}

\subsubsection{SNR and Link Budget Calculation}
The received signal power \( S \) is computed using Friis transmission equation \cite{skolnik1980introduction}:
\begin{equation}\label{eq:Friis}
    S = \frac{P_{t}G_{t}G_{r}\lambda^2}{(4\pi R)^2}
\end{equation}
where:
\begin{itemize}
    \item \( P_t = 20 \) W is the spacecraft transmission power.
    \item \( G_t \) is the transmit antenna gain, given by:
    \begin{equation}\label{eq:G_t}
        G_t = \frac{4\pi A}{\lambda^2}, \quad A = \pi \left(\frac{D}{2}\right)^2.
    \end{equation}
    For a spacecraft parabolic antenna of approximately 0.8 to 1 meter in diameter, this results in:
    \begin{equation}\label{eq:G_t_value}
        G_t \approx 47~\text{dBi}.
    \end{equation}
    \item \( G_r \) is the receive antenna gain, assuming a 5-meter parabolic dish onboard each GEO satellite:
    \begin{equation}\label{eq:G_r_value}
        G_r \approx 60-64~\text{dBi}.
    \end{equation}
    \item \( R = 1.0 \times 10^{11} \) m (0.7 AU).
    \item \( \lambda = \frac{c}{f} = 9.4 \times 10^{-3} \) m (Ka-band, 32 GHz).
\end{itemize}

Substituting into equation \ref{eq:Friis}:
\begin{equation}\label{eq:S_value}
    S = \frac{(20)(10^{4.7})(10^{6.2})(9.4 \times 10^{-3})^2}{(4\pi (1.0 \times 10^{11}))^2}
\end{equation}
\begin{equation}\label{eq:S_dBW}
    S \approx 1.13 \times 10^{-20} \text{ W} \quad \text{or} \quad -169.5 \text{ dBW}.
\end{equation}

\par\vspace{2\baselineskip}

Using the simplified logarithmic form of equation~\ref{eq:SNR_simple}, the single-sample signal-to-noise ratio (SNR) in dB units is calculated as:

\begin{equation}\label{eq:SNR_dB}
   \text{SNR}_{\text{eff}} = S_{\text{dB}} - N_{\text{T,dB}}.
\end{equation}

For the case of a 0.7 AU deep-space link, the received signal power and thermal noise power are given by:

\begin{equation}\label{eq:S_value_corrected}
   S_{\text{dB}} = -169.5~\text{dBW}, \quad N_{\text{T,dB}} = -174~\text{dBW}.
\end{equation}

Substituting these values into equation~\ref{eq:SNR_dB}, the effective single-sample SNR is:

\begin{equation}\label{eq:SNR_final_corrected}
   \text{SNR}_{\text{eff}} = (-169.5) - (-174) = 4.5~\text{dB}.
\end{equation}

The key link budget parameters used for this calculation are summarized in Table~\ref{tab:linkbudget}.

\begin{table}[hbt!]
    \centering
    \caption{Link budget parameters for a 0.7 AU deep-space mission.}
    \label{tab:linkbudget}
    \begin{tabular}{|l|c|c|l|}
    \hline
    \textbf{Parameter} & \textbf{Symbol} & \textbf{Value} & \textbf{Unit} \\
    \hline
    Distance & \(R\) & \(1.0\times10^{11}\) & m \\
    Transmit antenna diameter & \(D_t\) & 0.8--1 & m \\
    Transmit antenna gain & \(G_t\) & 47 & dBi \\
    Receive antenna diameter & \(D_r\) & 5 & m \\
    Receive antenna gain & \(G_r\) & 60--64 & dBi \\
    System noise temperature & \(T_{op}\) & 292.7 & K \\
    Receiver bandwidth & \(BW\) & \(1.0\times10^{3}\) & Hz \\
    Thermal noise power & \(N_{\text{T}}\) & -174 & dBW \\
    Received signal power & \(S\) & -169.5 & dBW \\
    Phase noise variance & \(\sigma^2\) & \(1.0\times10^{-2}\) & rad\(^2\) \\
    Effective SNR & \(\text{SNR}_{\text{eff}}\) & 4.5 & dB \\
    \hline
    \end{tabular}
\end{table}

\subsubsection{Aggregated SNR Effect on Phase Difference Estimation}

While the single-sample SNR defines the instantaneous measurement quality, the aggregate accuracy improves with coherent integration over multiple samples. For $N$ independent and Gaussian distributed phase samples, a lower bound for the variance of an unbiased phase estimator is given by the Cramér–Rao Lower Bound (CRLB). For long-baseline, dual-station interferometry, assuming Gaussian noise conditions (valid under thermal noise dominance), the CRLB for phase difference estimation is expressed as \cite{pourhomayoun2012cramer}:

\begin{equation} \label{CRLB}
\text{Var}(\Delta\phi) \geq  CRLB = \frac{1}{N \cdot \text{SNR}},
\end{equation}

where $\text{Var}(\Delta\phi)$ is the phase difference error variance, $N$ is the number of coherently integrated samples, and $\text{SNR}$ is the single-sample signal-to-noise ratio (in linear scale).

The corresponding standard deviation (1$\sigma$ error) in radians is:

\begin{equation}
    \sigma_{\Delta\phi} = \sqrt{\frac{1}{N \cdot \text{SNR}}}.
\end{equation}

\paragraph{Numerical Estimation for RINGS:}
Assuming a sampling interval of $1$ millisecond ($T_s = 1~\text{ms}$), and an integration time of $T_{\text{int}} = 120$ to $300$ seconds (2--5 minutes), the number of samples is:

\begin{equation}
    N = \frac{T_{\text{int}}}{T_s} = \frac{120}{0.001} \text{ to } \frac{300}{0.001} \quad \Rightarrow \quad N = 120{,}000 \text{ to } 300{,}000.
\end{equation}

Given the single-sample SNR of $4.5~\text{dB}$, corresponding to:

\begin{equation}
    \text{SNR}_{\text{linear}} = 10^{4.5/10} \approx 2.82,
\end{equation}

the CRLB-based phase standard deviation becomes:

\begin{itemize}
    \item For $N = 120{,}000$ (2 minutes):
    \begin{equation}
        \sigma_{\Delta\phi} = \sqrt{\frac{1}{120{,}000 \times 2.82}} \approx 0.0054~\text{rad} \quad (\sim 0.31^\circ)
    \end{equation}
    \item For $N = 300{,}000$ (5 minutes):
    \begin{equation}
        \sigma_{\Delta\phi} = \sqrt{\frac{1}{300{,}000 \times 2.82}} \approx 0.0034~\text{rad} \quad (\sim 0.20^\circ)
    \end{equation}
\end{itemize}

These results quantify the \emph{minimum achievable phase error} due to SNR-limited thermal noise under Gaussian assumptions. Even with a modest single-sample SNR of 4.5~dB, the accumulation of $10^5$--$3 \times 10^5$ samples through coherent integration reduces the phase error to below $0.3^\circ$, demonstrating the critical role of long integration times in enhancing measurement precision. This represents only the noise-limited contribution; additional error sources will be addressed in the following sections.

\subsection{Doppler Effect in GEO-Based VLBI}

Geostationary Earth Orbit (GEO) satellites move at approximately $V_{\text{GEO}} = 3075\,\text{m/s}$ in an Earth-Centered Inertial (ECI) frame, introducing significant Doppler shifts compared to terrestrial VLBI stations. In space-based Very Long Baseline Interferometry (VLBI), this Doppler effect must be carefully modeled, as it contributes a systematic bias to the measured phase.

The general expression for the dual-frequency differential phase, derived previously for a generic two-station configuration (see equation \ref{DDphi_dop_vector}), is recalled here for reference:
\begin{equation*}
    \Delta\Delta\phi = \frac{2\pi \Delta f_c}{c} \left[ B \cos\theta + \frac{r}{c} \Delta V_r \right]
\end{equation*}
where $B$ is the baseline length, $\theta$ is the angle between the baseline and the line-of-sight to the target, $r$ is a nominal range to the spacecraft, and $\Delta V_r = v_{r_1} - v_{r_2}$ is the differential radial velocity between the two stations.

In a symmetric GEO-based configuration, the two satellites are separated by $180^\circ$ in longitude. Their radial velocities relative to the target are:
\begin{equation}
    v_{r_1} = V_{\text{GEO}} \sin\theta, \quad v_{r_2} = -V_{\text{GEO}} \sin\theta
\end{equation}
so that:
\begin{equation}
    \Delta V_r = v_{r_1} - v_{r_2} = 2 V_{\text{GEO}} \sin\theta
\end{equation}

In this configuration, the differential range satisfies:
\begin{equation}
    r_1(t) - r_2(t) = B \cos\theta,
\end{equation}
but this applies only to the range difference, not to the individual ranges.

Substituting the expressions for $\Delta V_r$ and geometry into the general phase equation yields:
\begin{equation}
    \Delta\Delta\phi = \frac{2\pi \Delta f_c}{c} \left[ B \cos\theta + \frac{r}{c} \cdot 2 V_{\text{GEO}} \sin\theta \right] = \Delta\Delta\phi_{\text{static}} + \Delta\phi_{\text{error}}
\end{equation}
where the second term represents the Doppler-induced phase error accumulated over time.

To analyze this time-dependent component, we begin from the instantaneous Doppler frequency:
\begin{equation}
    \Delta f_{\text{Doppler}} = \frac{\Delta f_c}{c} \cdot \Delta V_r = \frac{2 \Delta f_c V_{\text{GEO}} \sin\theta}{c}
\end{equation}

The corresponding instantaneous phase rate is:
\begin{equation}
    \frac{d\phi_{\text{error}}}{dt} = 2\pi \Delta f_{\text{Doppler}} = \frac{4\pi \Delta f_c V_{\text{GEO}} \sin\theta}{c}
\end{equation}

Integrating over the interferometric integration time $T$ gives the accumulated phase error:
\begin{equation}
    \Delta\phi_{\text{error}} = \int_0^T \frac{d\phi_{\text{error}}}{dt} \, dt = \frac{4\pi \Delta f_c V_{\text{GEO}} \sin\theta}{c} \cdot T
\end{equation}

This expression defines the Doppler-induced phase drift over time for a dual-GEO VLBI system. It is linearly proportional to the radial velocity difference and integration duration, and thus geometry dependent.

We evaluate $\Delta\phi_{\text{error}}$ for three representative angles using the following parameters:
\[
\Delta f_c = 10^5 \, \text{Hz}, \quad V_{\text{GEO}} = 3075 \, \text{m/s}, \quad T = 300 \, \text{s}, \quad c = 3 \times 10^8 \, \text{m/s}.
\]

\begin{table}[h!]
\centering
\caption{Doppler-Induced Phase Error at Various Target Angles}
\begin{tabular}{|c|c|c|c|}
\hline
$\theta$ (deg) & $\Delta V_r$ [m/s] & $\Delta\phi_{\mathrm{error}}$ [rad] & Cycles \\
\hline
5  & 536.0  & 44.90  & 7.15 \\
45 & 4348.7 & 364.32 & 57.98 \\
85 & 6126.6 & 513.26 & 81.69 \\
\hline
\end{tabular}
\end{table}

As seen, the Doppler-induced phase error increases monotonically with angle $\theta$, reaching its maximum near $85^\circ$ where the relative radial velocity is largest. At this angle, the accumulated phase exceeds $500$ radians (more than 80 cycles). At $\theta = 45^\circ$, the Doppler contribution remains substantial, corresponding to approximately 364 radians (58 cycles). Even at small look angles (e.g., $\theta = 5^\circ$), the effect is non-negligible for VLBI systems, resulting in about 45 radians (7 cycles) of phase error.

This analysis shows that Doppler contributions in GEO-based VLBI are strongly geometry dependent and must be precisely modeled. Even in symmetric configurations, the velocity-induced phase drift may dominate the error budget. Consequently, resolving the inherent $2\pi$ phase ambiguity - through phase tracking, unwrapping, or a priori modeling - is essential to preserve measurement integrity.

\medskip

\noindent
Once corrected, the residual phase error can be interpreted as a small angular offset. The corresponding angular deviation is approximately:
\begin{equation}
    \Delta\theta \approx \frac{\lambda}{2\pi B \cos\theta} \cdot \Delta\phi_{\text{error}}
\end{equation}
where $\lambda$ is the signal wavelength, $B$ is the interferometric baseline, and $\theta$ is the nominal angle about which variations occur. This relation will be used in later sections to translate phase noise or post-correction residuals into angular uncertainty bounds.

\subsection{Time Synchronization and Clock Instability in GEO-Based VLBI}

Precise time synchronization between the two GEO satellites is a fundamental requirement in the RINGS system, as interferometric phase measurements are highly sensitive to clock offsets. Unlike terrestrial VLBI, which utilizes ultra-stable ground-based hydrogen masers with timing stability below $10^{-14}$ over 1000 seconds, spaceborne platforms must rely on smaller, more practical clocks compatible with space constraints. This introduces unique challenges due to limitations in size, weight, power consumption, and the inability to maintain continuous line-of-sight between the satellites for direct synchronization.

\subsubsection{Sources of Timing Error}

The total timing uncertainty in the RINGS system stems from two main factors:

\begin{itemize}
    \item \textbf{Clock Instability:} GEO satellites are typically equipped with Rubidium clocks, offering an Allan deviation of approximately $10^{-12}$ over 1000 seconds. This corresponds to a drift of 3.6 nanoseconds over one hour, equivalent to roughly 725 radians of phase error at 32 GHz.
    \item \textbf{Time Transfer Limitations:} Without continuous synchronization, onboard clock drift accumulates over time. Conventional methods like GNSS-based corrections mitigate long-term drift but leave residual errors between updates.
\end{itemize}

We conservatively estimate a post-calibrated timing uncertainty of approximately 100 ps, resulting in:
\begin{equation}
    \Delta\phi_{\text{clk}} = 2\pi f_0 \Delta t_{\text{clk}} \approx 2\pi \cdot 32 \times 10^9 \cdot 10^{-10} \approx 20~\text{rad},
\end{equation}
which corresponds to nearly 3 phase cycles. This level of error must be corrected or unwrapped during phase processing to avoid degradation of angular resolution.

\subsubsection{Synchronization Strategies}

Multiple approaches are considered for time synchronization in the RINGS architecture:

\paragraph{(1) Autonomous Direct Synchronization (Line-of-Sight Required)}

This method involves real-time bidirectional communication between the two GEO satellites:
\begin{itemize}
    \item \textbf{One-way time transfer:} A master satellite transmits a timing reference to synchronize the secondary node.
    \item \textbf{Two-way time transfer:} Bidirectional exchange of timing signals allows for real-time drift estimation and compensation.
\end{itemize}

However, in the RINGS configuration with satellites separated by $180^\circ$ in longitude, Earth blocks direct line-of-sight, rendering this option infeasible without an additional relay satellite positioned at an intermediate orbital slot.

\paragraph{(2) Hybrid Internal-External Synchronization}

A more practical solution involves combining stable onboard clocks with periodic corrections using external references:

\begin{itemize}
    \item \textbf{GNSS-based drift correction:} Each GEO satellite receives GNSS signals independently to periodically calibrate its internal clock. This method maintains long-term stability but does not resolve real-time relative drift during interferometric sessions.
    \item \textbf{Post-processing correlation:} Residual clock offsets are estimated and removed during ground-based data processing. This technique corrects for slow drifts but requires careful modeling to avoid loss of phase coherence.
\end{itemize}

\subsubsection{Potential Improvements with Higher-Precision Spaceborne Clocks}

To further reduce timing-induced phase errors, the use of space-qualified Cesium clocks can be considered. Cesium standards offer Allan deviations on the order of $10^{-13}$, representing a tenfold improvement over Rubidium.

For example, a Cesium clock with $10^{-13}$ drift over one hour yields:
\begin{equation}
    \Delta t_{\text{Cesium}} = 10^{-13} \times 3600~\text{s} = 0.36~\text{ns},
\end{equation}
leading to a corresponding phase error:
\begin{equation}
    \Delta\phi_{\text{Cesium}} = 2\pi f_0 \Delta t_{\text{Cesium}} \approx 2\pi \cdot 32 \times 10^9 \cdot 0.36 \times 10^{-9} \approx 72.5~\text{rad},
\end{equation}
which is approximately 11.5 cycles.

With post-processing and GNSS correction, it is reasonable to assume residual errors can be reduced to around 30 ps, yielding:
\begin{equation}
    \Delta\phi_{\text{Cesium-residual}} = 2\pi \cdot 32 \times 10^9 \cdot 30 \times 10^{-12} \approx 6~\text{rad}.
\end{equation}

This represents a significant improvement over the Rubidium case, reducing phase uncertainty by approximately a factor of three.

\subsubsection{Inter-Satellite Synchronization Links}

Another promising method is to implement inter-satellite synchronization links using dedicated "over-the-air" communications. Although the current RINGS geometry prevents direct line-of-sight, potential alternatives include:

\begin{itemize}
    \item \textbf{Relay via third GEO satellite:} Placing a third satellite at an intermediate longitude could facilitate continuous time transfer.
    \item \textbf{Optical or RF cross-links:} Advanced inter-satellite links could provide real-time time transfer with precision approaching tens of picoseconds, using two-way methods to cancel propagation delays.
\end{itemize}

By combining high-stability onboard clocks with periodic GNSS calibration, post-correlation correction, and potentially inter-satellite synchronization links, the RINGS system can reduce timing-induced phase errors to acceptable levels for high-accuracy VLBI operations.

\subsection{Satellite Orbit Determination (OD) Accuracy}

A key differentiator between terrestrial and satellite-based VLBI lies in the accuracy with which the station positions are known. Ground VLBI antennas are surveyed with centimeter-level precision, whereas the positions of GEO satellites are typically known to 10–100 meters unless dedicated tracking systems such as GPS, Satellite Laser Ranging (SLR), or inter-satellite links are employed.

In conventional GEO operations, a 10–100 meter position error is common and would impose a significant limitation on VLBI performance. The uncertainty in the baseline vector $B$ directly translates into angular measurement error via the linearized relation:

\begin{equation}
    \delta \theta \approx \frac{\delta B}{B} \cdot \theta,
\end{equation}

For example, assuming a 10-meter orbit determination error over an $80{,}000$ km GEO-GEO baseline:

\begin{equation}
    \frac{10}{80{,}000{,}000} = 1.25 \times 10^{-7} \quad \Rightarrow \quad \delta \theta \approx 125~\text{prad}~\text{for}~\theta = 1~\text{mrad}.
\end{equation}

This level of error is unacceptably large for precision deep-space VLBI applications.

However, since the RINGS system is specifically designed for VLBI navigation, and since the GEO satellite locations are needed \emph{a posteriori} (at the time of measurement correlation rather than in real time), we assume a dedicated orbit determination process is employed. This includes precise GNSS-based tracking, laser ranging when applicable, and fusion with inter-satellite link data. As a result, we conservatively assume that a post-processed orbital knowledge of \textbf{0.5 meters (50 cm)} is achievable.

For a 0.5-meter OD error, the resulting angular error is:

\begin{equation}
    \frac{0.5}{80{,}000{,}000} = 6.25 \times 10^{-9} \quad \Rightarrow \quad \delta \theta \approx 6.25~\text{nrad}~\text{for}~\theta = 1~\text{mrad}.
\end{equation}

This level of station location uncertainty, while still larger than for ground VLBI, is consistent with the overall system error budget and can be mitigated through post-correlation analysis. It represents the value adopted for the final performance estimation in Table~\ref{tab:vlbi_comparison}.

\subsection{Elimination of Atmospheric Path Errors}

An important advantage of the proposed RINGS architecture is the complete elimination of tropospheric and ionospheric phase delay errors. In terrestrial VLBI, these effects constitute significant components of the total error budget:
\begin{itemize}
    \item Tropospheric wet delay: up to 20--30 ps RMS
    \item Ionospheric group delay: up to 15--20 ps, especially at X-band
    \item Solar plasma effects: variable, geometry dependent
\end{itemize}

Since the GEO satellites operate entirely above the Earth’s atmosphere, these contributions vanish. This removes the need for auxiliary calibration systems such as microwave radiometers, water vapor profilers, or dual-frequency plasma compensation links.

Thus, while the RINGS system introduces new error sources (e.g., clock and OD uncertainty), it also removes several dominant ones from the classical ground-based budget. This tradeoff is a key feature of the proposed space-based VLBI paradigm.

\subsection{Error Budget Summary and Analysis}

Table~\ref{tab:vlbi_comparison} summarizes the error contributions for both traditional terrestrial VLBI and the proposed GEO-based VLBI configuration (RINGS). We adopt the same error budget methodology used for ground-based VLBI and Delta-DOR systems, as extensively detailed in \cite{becker2013autonomous, curkendall2013delta,book2013delta, fiori2022deep,iess2014astra}. This ensures a direct and consistent comparison between terrestrial and space-based architectures.

For the terrestrial case, the Root Sum Square (RSS) timing error is approximately 0.063 nanoseconds, corresponding to an angular uncertainty of 2.39 nanoradians. In contrast, the GEO-based RINGS system yields an RSS of 1.044 nanoseconds, resulting in an angular uncertainty of approximately 3.73 nanoradians.

The dominant contributors to the satellite-based VLBI error budget are as follows:

\begin{itemize}
    \item \textbf{Spacecraft SNR contribution:} Derived from the Cramér-Rao Lower Bound (CRLB) for phase estimation under Gaussian noise conditions. For a 2 to 5-minute integration time and a sampling interval of 1 millisecond, the accumulated phase error is between $0.0054$ and $0.0034$ radians. Converting this to timing uncertainty at Ka-band ($f_0 = 32$~GHz) yields:
    \[
    \Delta t_{\text{SNR}} = \frac{\Delta\phi}{2\pi f_0}
    \]
    Using the worst-case (shorter integration time), this results in:
    \[
    \Delta t_{\text{SNR}} \approx \frac{0.0054}{2\pi \times 32 \times 10^9} \approx 27~\text{ps}
    \]
    For conservatism, we round to $\Delta t_{\text{SNR}} \approx 30~\text{ps}$.
    
    \item \textbf{Clock instability:} The use of space-qualified Rubidium clocks onboard the GEO satellites introduces residual timing errors after GNSS-based correction and post-correlation calibration. The residual clock error is estimated at approximately 300 picoseconds, corresponding to a phase error of about 20 radians at 32 GHz.
    
    \item \textbf{Station location uncertainty:} Unlike terrestrial VLBI antennas, whose positions are known to centimeter accuracy, satellite-based systems suffer from orbit determination limitations. While standard GEO orbit knowledge is typically in the 10--100 meter range, the RINGS system assumes dedicated post-processed orbit determination achieving 0.5 meter accuracy. For an $80{,}000$ km baseline, this leads to:
    \[
    \frac{0.5}{80{,}000{,}000} = 6.25 \times 10^{-9}
    \]
    Translating this into timing error over light speed yields:
    \[
    \delta t_{\text{OD}} \approx 1~\text{nsec}
    \]
    This is currently the dominant contributor to the overall error budget.
\end{itemize}

Several error sources present in terrestrial VLBI - such as tropospheric fluctuations, ionospheric delays, and quasar coordinate uncertainties - are effectively eliminated or negligible in the RINGS system, since the receivers operate in space and directly observe the spacecraft signal.

While the absolute Root Sum Square (RSS) error is higher in the GEO-based scenario, the system still meets the angular accuracy requirements for deep-space navigation. Moreover, the inherent advantages of space-based VLBI—including continuous target visibility, immunity to atmospheric effects, and extended baselines—help compensate for much of the increased error budget.

Future improvements may include the deployment of higher-precision onboard clocks (e.g., space-qualified cesium standards), inter-satellite synchronization links, and further enhancements in orbit determination accuracy.

\begin{table}[hbt!]
    \centering
    \caption{Comparison of Terrestrial VLBI and Satellite-Based VLBI Error Budgets}
    \begin{tabular}{|l|c|c|}
        \hline
        \textbf{Error Source} & \textbf{Terrestrial VLBI (nsec)} & \textbf{Satellite VLBI (nsec)} \\
        \hline
        Quasar-SNR (QSNR) & 0.033 & - \\
        Spacecraft SNR & 0.019 & 0.030 \\
        Clock Instability & 0.004 & 0.300 \\
        Dispersive Phase & 0.030 & 0.030 \\
        Station Location Uncertainty & 0.010 & 1.000 \\
        Earth Orientation & 0.013 & 0.013 \\
        Troposphere Systematic (Zenith) & 0.012 & - \\
        Troposphere Fluctuations & 0.019 & - \\
        Ionospheric Effects & 0.019 & - \\
        Solar Plasma Effects & 0.006 & 0.006 \\
        Quasar Coordinates & 0.020 & - \\
        \hline
        \textbf{Root Sum Square (RSS)} & \textbf{0.063} & \textbf{1.044} \\
        \hline
        \textbf{Angular Error (nano-radians)} & \textbf{2.39} & \textbf{3.73} \\
        \hline
    \end{tabular}
    \label{tab:vlbi_comparison}
\end{table}

\noindent
According to this analysis, the GEO-based VLBI system (RINGS) exhibits a higher overall Root Sum Square (RSS) error compared to traditional terrestrial VLBI, primarily due to clock instability and station location uncertainty. The calculated satellite-based RSS timing error is 1.044~nsec, corresponding to an angular uncertainty of 3.73~nanoradians compared to the terrestrial VLBI error of 0.063~nsec and 2.39~nrad respectively.

However, the space-based system still maintains error levels within the same order of magnitude as terrestrial VLBI. If spacecraft-based VLBI is combined with periodic quasar calibration, further improvements in angular resolution could be achieved, particularly in correcting baseline and clock biases.

Although the GEO-based VLBI configuration exhibits a slightly higher error floor, its additional advantages—including continuous observation windows, significantly reduced effects of atmospheric noise, and the potential for higher Sun-Earth-Probe (SEP) angles - strongly justify continued development and field experimentation. The proposed method offers a promising alternative to Earth-based VLBI, especially in scenarios where terrestrial station geometry limits availability or accuracy.

\section{Simulation and Modeling} \label{simulation}

To validate the proposed RINGS concept, a dedicated simulation framework was developed combining the Ansys/AGI System Tool Kit (STK) and MATLAB. The goal is to assess system performance under realistic orbital conditions and quantify both angular measurement accuracy and system availability.

\subsection{Simulation Structure}

\begin{figure}[hbt!]
\centering
\includegraphics[width=0.99\linewidth, height=7cm]{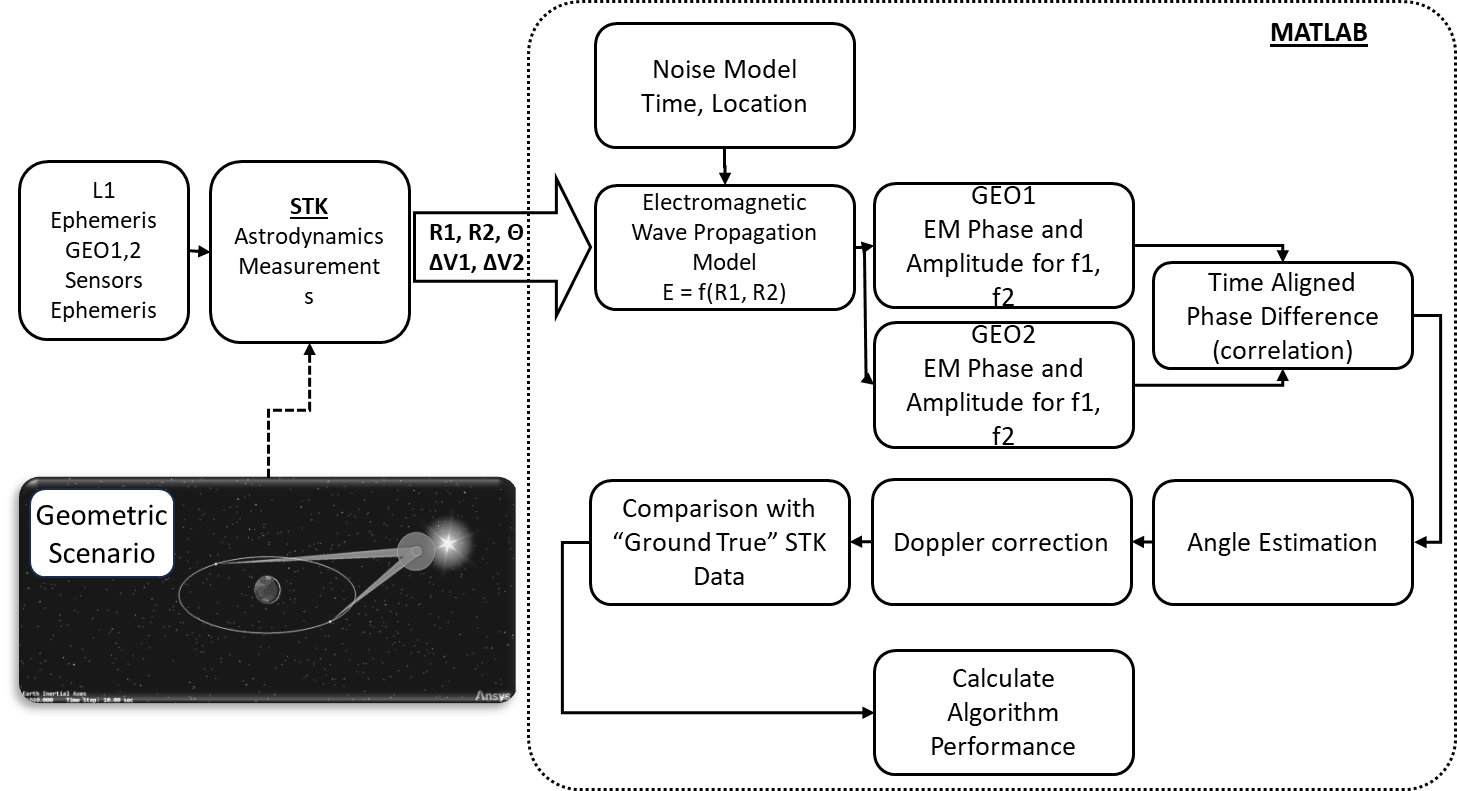}
\caption{Block diagram of the STK-MATLAB simulation for the RINGS system.}
\label{fig: RINGS simulation STK}
\end{figure}

The simulation consists of two integrated components:

\begin{itemize}
    \item \textbf{STK Orbital Scenario:} A representative deep space mission geometry was modeled using the Sun–Earth Lagrangian point L1 (SEL1) as an example target. Two GEO satellites were placed in dual longitude slots, creating a long interferometric baseline. STK provides the orbital dynamics, including relative positions, velocities, and baseline-spacecraft angles ($\theta$). While SEL1 is used as a reference case, the geometric considerations and availability analysis are generally applicable to any deep space target located in the ecliptic plane.
    
    \item \textbf{MATLAB Interferometric Model:} A custom MATLAB code emulates the full radiometric interferometry process. The model includes electromagnetic wave propagation from the spacecraft, phase sampling at the GEO receivers, dual-frequency processing, and cross-correlation to extract the differential phase. The simulation structure is illustrated in Fig.~\ref{fig: RINGS simulation STK}.
\end{itemize}

The simulation accounts for Doppler correction, phase ambiguity handling, and basic noise sources such as clock biases and J2 perturbations. At this stage, perfect time synchronization was assumed.

\subsection{Preliminary Angular Estimation Results}

The interferometric measurement process was tested for two representative baseline-spacecraft angles:

\begin{itemize}
    \item $\theta \approx 22\degree$ (short baseline projection, elevation-sensitive geometry)
    \item $\theta \approx 72\degree$ (long baseline projection, azimuth-sensitive geometry)
\end{itemize}

Each measurement session lasted 120 seconds, using two sampling rates: $10~\text{ms}$ (100 Hz) and $1~\text{ms}$ (1 kHz). A carrier frequency of $3~\text{GHz}$ (S-band) was selected for preliminary analysis.

\begin{figure}[hbt!]
\centering
\includegraphics[width=20pc]{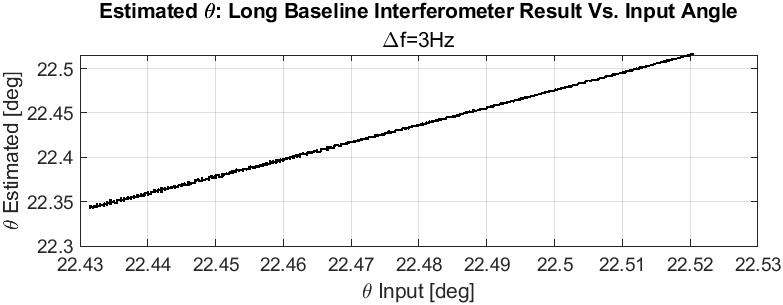}
\caption{Simulation results for interferometric angle estimation at $\theta\approx22\degree$.}
\label{fig: Theta estimate 22}
\end{figure}

\begin{figure}[hbt!]
\centering
\includegraphics[width=20pc]{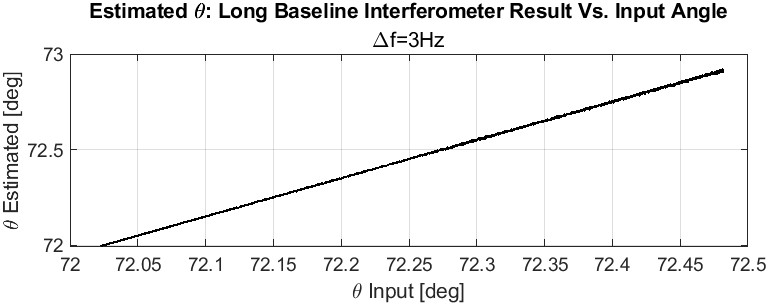}
\caption{Simulation results for interferometric angle estimation at $\theta\approx72\degree$.}
\label{fig: Theta estimate 72}
\end{figure}

Simulation results are presented in Fig.~\ref{fig: Theta estimate 22} and Fig.~\ref{fig: Theta estimate 72}. The RINGS algorithm successfully estimates the angle of arrival $\theta$, with residual errors primarily due to phase ambiguity resolution challenges—an inherent difficulty in long-baseline VLBI systems. These results reflect raw, unfiltered measurements without sequential estimation or smoothing.

\subsection{System Availability Simulation}

Beyond angular estimation, the simulation also analyzed system availability compared to terrestrial VLBI arrays. Here, "availability" refers to the geometric condition of simultaneous line-of-sight (LOS) from both stations to the spacecraft—not system reliability or hardware uptime.

Two GEO satellites were positioned at $100\degree$ West and $80\degree$ East, creating a $180\degree$ orbital separation. For terrestrial comparison, the Deep Space Network (DSN) stations in Goldstone, Canberra, and Madrid were modeled.

A full-year simulation for 2030 was performed in STK, recording LOS conditions every 10 minutes. Although SEL1 was used as the target, similar availability results are expected for any deep space mission in the ecliptic plane due to the geometry of GEO coverage.

\begin{figure}[hbt!]
\centering
\includegraphics[width=0.99\linewidth]{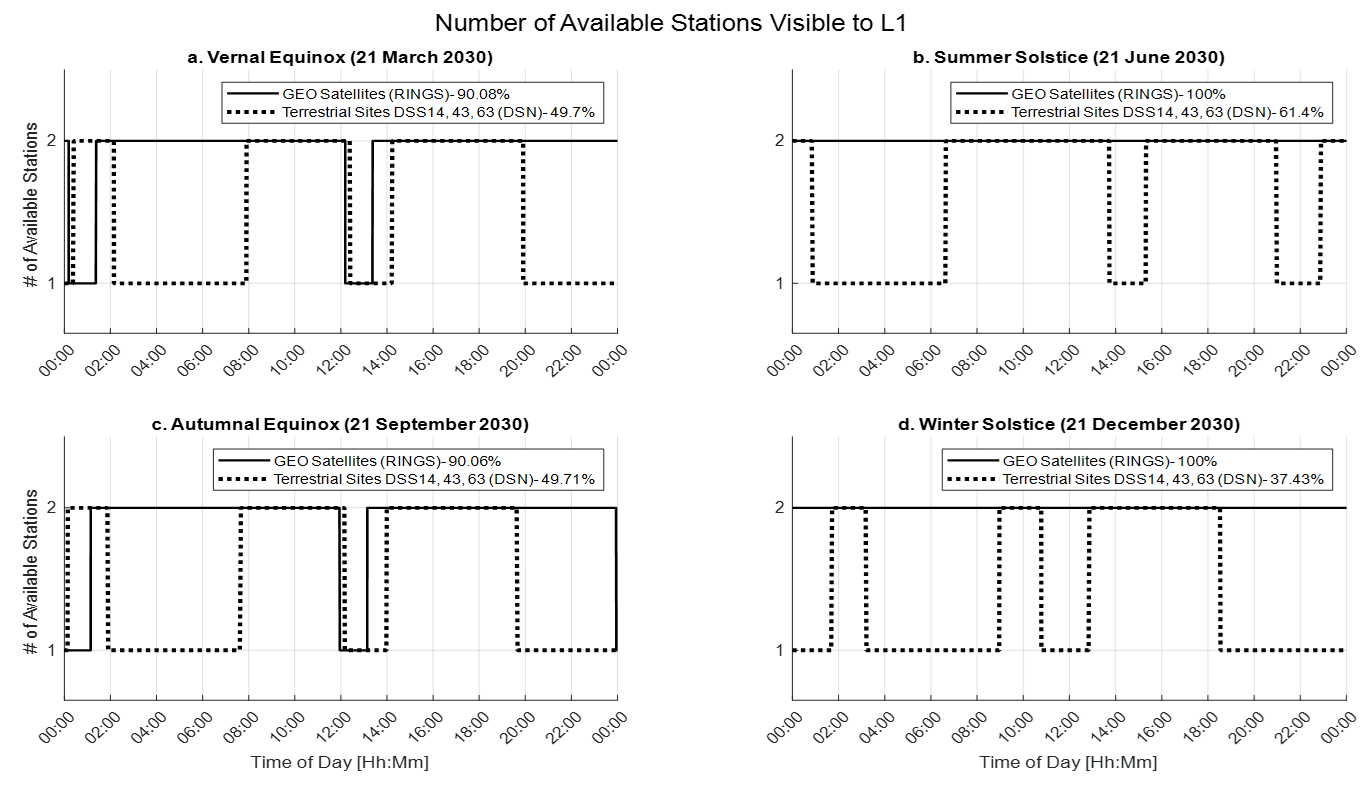}
\caption{Number of sites visible to deep space reference location (SEL1) vs. time, over 24 hours. GEO-based VLBI nearly doubles the interferometry availability compared to terrestrial systems.}
\label{fig: System availability}
\end{figure}

\subsubsection{Availability Results}

\begin{itemize}
    \item \textbf{Terrestrial VLBI:} Achieves approximately 49.66\% availability due to Earth rotation and geographic constraints. Typical daily operation involves three 4-hour overlap episodes between DSN stations.
    
    \item \textbf{GEO-based VLBI:} Provides 98\% availability, with only short interruptions during GEO eclipses near equinoxes. For certain targets such as SEL1, the system also offers a favorable Sun–Earth–Probe (SEP) angle, reducing solar proximity effects on RF measurements.
\end{itemize}

This substantial improvement in geometric availability translates to increased measurement duty cycles, reduced operational constraints, and enhanced support for continuous deep-space navigation. It represents one of the principal operational advantages of the RINGS concept over traditional ground-based VLBI.

\section{Conclusions}\label{conclusions}

Radiometric interferometry is a fundamental tool for deep space navigation and spacecraft orbit determination. In this paper, we proposed and analyzed a system concept in which two geostationary satellites serve as an interferometric baseline to enable continuous angular tracking of deep space targets. This GEO-based Very Long Baseline Interferometry (VLBI) configuration, termed RINGS, introduces a new architecture that extends the capabilities of existing ground-based VLBI systems.

Using a structured error budget analysis consistent with NASA and ESA methodologies for VLBI and Delta-DOR systems, we evaluated the dominant error sources in the RINGS system. The total Root Sum Square (RSS) timing error was found to be $1.044$ nanoseconds, corresponding to an angular uncertainty of $3.73$ nanoradians. For comparison, the equivalent RSS error in terrestrial VLBI systems is approximately $0.063$ nanoseconds (or $2.39$ nanoradians). Despite this increase, the RINGS system remains within the same order of magnitude, which is a notable result given the unique advantages of the space-based configuration.

Several contributing factors were analyzed in detail:
\begin{itemize}
    \item \textbf{Signal-to-noise ratio (SNR)}: For a 2 to 5-minute integration time with a $1$~kHz sampling rate, the phase error due to SNR was computed to be approximately $0.0054$ radians, resulting in a timing uncertainty of about $30$ picoseconds.
    \item \textbf{Clock instability}: Post-GNSS correction and post-processing calibration leave a residual clock error estimated at $300$ picoseconds, corresponding to $\sim20$ radians of phase uncertainty at Ka-band.
    \item \textbf{Orbit determination (OD) uncertainty}: Assuming a dedicated post-processed OD solution with $0.5$ meter baseline knowledge, the resulting error contribution is $1$ nanosecond, dominating the baseline uncertainty term.
\end{itemize}

In addition to the quantitative error analysis, availability simulations demonstrated a significant operational advantage. The GEO-based system achieves approximately $98\%$ time availability for interferometric observations of SEL1, compared to $49.66\%$ for traditional terrestrial VLBI sites. This enhanced visibility opens new opportunities for continuous tracking and improved duty cycles, particularly for missions where Earth-based line-of-sight is periodically unavailable.

This work demonstrates that GEO-based VLBI is both theoretically feasible and practically promising for deep space navigation. While the initial error budget indicates a slightly higher absolute error compared to ground-based VLBI, the system remains within the same performance order, with substantial potential for improvement. Future enhancements - such as higher-precision space-borne clocks, dedicated inter-satellite synchronization links, and advanced post-processing techniques - can further reduce the error contribution.

As the number of deep space and interplanetary missions continues to grow, the development of space-based navigation infrastructure becomes increasingly relevant. The RINGS concept offers a complementary approach to existing ground-based VLBI systems, reducing reliance on terrestrial stations and enabling continuous deep space tracking with extended availability and improved baseline geometry. Further research, extended simulation campaigns, and eventual field experimentation are warranted to refine this method and assess its operational potential for future space navigation architectures.

\section*{Acknowledgments}
This work was supported by the Peter Munk Research Institute (PMRI) at the Technion and by the Gordon Center for Systems Engineering 2024 grants. The authors also gratefully acknowledge receiving the Gemunder Prize for Space-Defense Related Technologies, which recognized this research in 2025.

We sincerely thank Dr. Alex Frid from the Asher Space Research Institute (ASRI) at the Technion, Haifa, for his valuable insights. We are also indebted to Almog Yanku, Daniel Shanan, and Itai Carmeli for their dedicated support in implementing the MATLAB and STK codes used to demonstrate key aspects of the proposed navigation concept.

\vspace{1em}
\noindent\textbf{Conflicts of Interest:} The authors declare no conflicts of interest.


\begin{thebibliography}{10}

\bibitem{turan2022autonomous}
E.~Turan, S.~Speretta, and E.~Gill, ``Autonomous navigation for deep space small satellites: Scientific and technological advances,'' {\em Acta Astronautica}~{\bf 193}, pp.~56--74, 2022.

\bibitem{vittorio2021autonomous}
F.~Vittorio, {\em Autonomous navigation for interplanetary CubeSats at different scales}.
\newblock PhD thesis, Politecnico di Milano, 2021.

\bibitem{ParkWonKwon2017}
Y.~Park, J.-H. Won, and K.-H. Kwon, ``Performance analysis of multi-constellation and multi-frequency gnss receivers in deep space,'' in {\em Proceedings of the 30th International Technical Meeting of the Satellite Division of The Institute of Navigation (ION GNSS+ 2017},  2017.

\bibitem{andreis2021overview}
E.~Andreis, V.~Franzese, F.~Topputo, {\em et~al.}, ``An overview of autonomous optical navigation for deep-space cubesats,'' in {\em International Astronautical Congress: IAC Proceedings},  pp.~1--11, 2021.

\bibitem{book2013delta}
G.~BOOK, ``Delta-dor—technical characteristics and performance,'' 2013.

\bibitem{curkendall2013delta}
D.~W. Curkendall and J.~S. Border, ``Delta-dor: The one-nanoradian navigation measurement system of the deep space network—history, architecture, and componentry,'' {\em The Interplanetary Network Progress Report}~{\bf 42}, p.~193, 2013.

\bibitem{ely2022comparison}
T.~Ely, S.~Bhaskaran, N.~Bradley, T.~J.~W. Lazio, and T.~Martin-Mur, ``Comparison of deep space navigation using optical imaging, pulsar time-of-arrival tracking, and/or radiometric tracking,'' {\em The Journal of the Astronautical Sciences}~{\bf 69}(2), pp.~385--472, 2022.

\bibitem{mudgway2001uplink}
D.~J. Mudgway, {\em Uplink-downlink: a history of the nasa deep space network, 1957-1997}, vol.~4227, National Aeronautics and Space Administration, 2001.

\bibitem{doat2018esa}
Y.~Doat, M.~Lanucara, P.-M. Besso, T.~Beck, G.~Lorenzo, and M.~Butkowic, ``Esa tracking network--a european asset,'' in {\em 2018 SpaceOps Conference},  p.~2306, 2018.

\bibitem{layland1997evolution}
J.~Layland and L.~Rauch, ``The evolution of technology in the deep space network: A history of the advanced systems program,'' {\em Jet Propulsion Laboratory, Pasadena, CA, TDA Prog. Rep}~{\bf 89}, 1997.

\bibitem{schuh2012vlbi}
H.~Schuh and D.~Behrend, ``Vlbi: A fascinating technique for geodesy and astrometry,'' {\em Journal of geodynamics}~{\bf 61}, pp.~68--80, 2012.

\bibitem{hellerschmied2018satellite}
A.~Hellerschmied, ``Satellite observations with vlbi,'' 2018.

\bibitem{fiori2022deep}
F.~Fiori, P.~Tortora, M.~Zannoni, A.~Ardito, M.~Menapace, G.~Bellei, F.~Budnik, T.~Morley, M.~Mercolino, and R.~Orosei, ``Deep space orbit determination via delta-dor using vlbi antennas,'' {\em CEAS Space Journal}~{\bf 14}(2), pp.~421--430, 2022.

\bibitem{iess2014astra}
L.~Iess, M.~Di~Benedetto, N.~James, M.~Mercolino, L.~Simone, and P.~Tortora, ``Astra: Interdisciplinary study on enhancement of the end-to-end accuracy for spacecraft tracking techniques,'' {\em Acta Astronautica}~{\bf 94}(2), pp.~699--707, 2014.

\bibitem{spencer2023interplanetary}
D.~B. Spencer and D.~Conte, {\em Interplanetary Astrodynamics}, CRC Press, 2023.

\bibitem{becker2013autonomous}
W.~Becker, M.~G. Bernhardt, and A.~Jessner, ``Autonomous spacecraft navigation with pulsars,'' {\em arXiv preprint arXiv:1305.4842} , 2013.

\bibitem{felli2012very}
M.~Felli and R.~E. Spencer, {\em Very Long Baseline Interferometry: Techniques and Applications}, vol.~283, Springer Science \& Business Media, 2012.

\bibitem{turner2012johnson}
C.~S. Turner, ``Johnson-nyquist noise,'' {\em url: http://www. claysturner. com/dsp/Johnson-NyquistNoise} , 2012.

\bibitem{skolnik1980introduction}
M.~I. Skolnik {\em et~al.}, {\em Introduction to radar systems}, vol.~3, McGraw-hill New York, 1980.

\bibitem{pourhomayoun2012cramer}
M.~Pourhomayoun and M.~Fowler, ``Cramer-rao lower bounds for estimation of phase in lbi based localization systems,'' in {\em 2012 Conference Record of the Forty Sixth Asilomar Conference on Signals, Systems and Computers (ASILOMAR)},  pp.~909--911, IEEE, 2012.

\end{thebibliography}
\end{document}